\documentclass{article}
\usepackage[utf8]{inputenc}
\usepackage{verbatim}
\usepackage{listings}
\usepackage[a4paper, margin=0.9in]{geometry}
\usepackage{braket}
\usepackage{enumitem, kantlipsum}
\usepackage{graphicx}
\usepackage[dvipsnames,table,xcdraw]{xcolor}
\usepackage{multicol}
\usepackage{mathtools}

\usepackage{csquotes}
\usepackage{amsmath}

\usepackage{url}

\usepackage{subfiles}
\usepackage{qcircuit}
\usepackage{amsthm,amssymb}
\usepackage{fdsymbol}
\usepackage{rotating}
\usepackage{caption}
\usepackage{hyperref}

\usepackage{xcolor}
\usepackage{sectsty}
\sectionfont{\color{purple}}
\subsectionfont{\color{teal}}
\subsubsectionfont{\color{olive}}

\usepackage{arydshln}

\usepackage[ruled,vlined]{algorithm2e}

\definecolor{codegreen}{rgb}{0,0.6,0}
\definecolor{codegray}{rgb}{0.5,0.5,0.5}
\definecolor{codepurple}{rgb}{0.58,0,0.82}
\definecolor{backcolour}{rgb}{0.95,0.95,0.92}
\lstdefinestyle{mystyle}{
    backgroundcolor=\color{backcolour},   
    commentstyle=\color{codegreen},
    keywordstyle=\color{magenta},
    numberstyle=\tiny\color{codegray},
    stringstyle=\color{codepurple},
    basicstyle=\ttfamily\footnotesize,
    breakatwhitespace=false,         
    breaklines=true,                 
    captionpos=b,                    
    keepspaces=true,                 
    numbers=left,                    
    numbersep=5pt,                  
    showspaces=false,                
    showstringspaces=false,
    showtabs=false,                  
    tabsize=2
}
\let\OLDthebibliography\thebibliography
\renewcommand\thebibliography[1]{
  \OLDthebibliography{#1}
  \setlength{\parskip}{0pt}
}

\title{\LARGE QKSA: Quantum Knowledge Seeking Agent\\
\large motivation, core thesis and baseline framework}

\usepackage[affil-it]{authblk}

\author{Aritra Sarkar}
\affil{Department of Quantum \& Computer Engineering,
\protect\\Faculty of Electrical Engineering, Mathematics and Computer Science,
\protect\\Delft University of Technology, Delft, The Netherlands}
\date{} 

\begin{document}

\maketitle





\section{Introduction} \label{s1}


In this article we present the motivation and the core thesis towards the implementation of a Quantum Knowledge Seeking Agent (QKSA).
QKSA is a general reinforcement learning agent that can be used to model classical and quantum dynamics.
It merges ideas from universal artificial general intelligence, constructor theory and genetic programming to build a robust and general framework for testing the capabilities of the agent in a variety of environments.
It takes the artificial life (or, animat) path to artificial general intelligence where a population of intelligent agents are instantiated to explore valid ways of modeling the perceptions.
The multiplicity and survivability of the agents are defined by the fitness, with respect to the explainability and predictability, of a resource-bounded computational model of the environment.
This general learning approach is then employed to model the physics of an environment based on subjective observer states of the agents.
A specific case of quantum process tomography as a general modeling principle is presented.
The various background ideas and a baseline formalism is discussed in this article which sets the groundwork for the implementations of the QKSA that are currently in active development.

Section~\ref{s2} presents a historic overview of the motivation behind this research
In Section~\ref{s3} we survey some general reinforcement learning models and bio-inspired computing techniques that forms a baseline for the design of the QKSA.
Section~\ref{s4} presents an overview of field of quantum artificial agents and the observer based operational theory that the QKSA aims to learn.
In Section~\ref{s5} and \ref{s6} we presents the salient features and a formal definition of our model.
In Section~\ref{s7}  we present the task of quantum process tomography (QPT) as a general task to test our framework.
In Section~\ref{s8} we conclude the discussion with suggestive future directions for implementing the QKSA.

\section{Historic overview} \label{s2}

\subsection{Scientific models}

The \textit{scientific method} can be traced back to philosophers from ancient civilizations, which over millennia has guided the empirical method of acquiring knowledge via careful observation and applying rigorous skepticism about what is observed, circumventing cognitive assumptions that can distort the interpretation. 
It involves formulating hypotheses based on such observations, via induction, and eventually experimental testing of deductions and refinement of the hypotheses.
This can be considered a simple prescription for general intelligence.

The formulated \textit{hypothesis} typically serves two purposes.
Firstly, the hypothesis is useful if it can be used for \textit{predicting} the outcomes of future repetition of the experiment.
There can be cases where this is not possible, e.g. models of the origin of the universe.
In such cases, the hypothesis should be able to independently reproduce the observed data.
However, prediction is typically not the only goal for forming a hypothesis.
As an example, the heliocentric and the geocentric model can both predict the motion of most objects in the sky and based on relativity, there is not preferred frame of reference.
The preferred model is based on the ease to semantically explain the model from the assumptions we already consider.
Thus, the hypothesis often serves as a \textit{model} of the process.
Occam's razor or the \textit{principle of parsimony} is used as an abductive heuristic in the development of theoretical models.
This philosophical razor advocates that when presented with competing hypotheses about the same prediction, one should select the solution with the fewest assumptions.
Since for any phenomena there might be an infinitely large number of valid explanations, this preference is based on the falsifiability criterion for more complex alternatives while experimental testing.
In mathematical systems, reducing the number of assumptions corresponds to devising a minimal \textit{system of axioms} that can be used to prove a set of statements.

Depending on the underlying logic, the problem of deciding the validity of a statement from the axioms may vary from trivial to impossible.
For example, in propositional logic, it is possible to design (i.e. it is decidable) an \textit{automatic theorem prover} (ATP) but might require exponential-time (co-NP) for it to execute.
For first order theories like Peano arithmetic, contrary to David Hilbert's vision of axiomatizing mathematics, Kurt Gödel's incompleteness theorem~\cite{godel1931formal} allows invalid formulas that cannot always be recognized as well as true but undecidable statements.
Thus, an ATP might fail to terminate while searching for a proof by \textit{deduction}.
Despite this theoretical limit, in practice, theorem provers can solve many hard problems.
Proof-based general intelligence models has been proposed by Jürgen Schmidhuber as the \textit{Gödel machine}~\cite{schmidhuber2007godel}, which eventually will find a proof of a self-improvement and evolve to optimally solve problems.

As opposed to the certainty of the argument from the premise in deduction, in \textit{inductive reasoning} the truth of the conclusion is inherently probable, based upon the evidence given.
The premises supplies some evidence, but not full assurance, of the truth of the conclusion.
While there has been philosophical discussions on the advantages and disadvantages of induction, it has been widely applied to many areas, especially in machine learning.
\textit{Bayesian inference} does not determine which beliefs are a priori rational, but rather determines how we should rationally change the beliefs we have when presented with evidence. 
In Bayesian inference, a prior probability is chosen for a hypothesis (based on previously acquired domain knowledge or logic) and then the strength of our belief in that hypothesis is adjusted in a precise manner according to Bayes' theorem when new evidence is acquired.

\subsection{Computational automata}

The precise definition of an algorithmic process was required to prove whether a statement is decidable or not from the set of axioms.
This led to the definition of the \textit{Turing machine} as a mathematical model of computation that manipulates symbols on a strip of tape according to a table of rules.
Alan Turing invented this model to prove properties of computation in general and the uncomputability of the Entscheidungsproblem~\cite{turing1936computable}.
Despite the model's simplicity, given an arbitrary computer algorithm, a Turing machine capable of simulating that algorithm's logic can be constructed and thus is a general example of a central processing unit (CPU).

A Turing machine that is able to simulate any other Turing machine is called a \textit{universal Turing machine} (UTM).
This \textit{universality} is shared between various other computing models like lambda calculus and cellular automata, and any computational procedure can be translated among these efficiently by incurring a polynomial time overhead.
This notion is captured as the \textit{Church-Turing thesis} which states that Turing machines can be used to define an effective methods in logical manipulation of an algorithm using a mechanical procedure.

\subsection{Algorithmic information}

Despite having a different original motivation, the UTM model eventually led to the development of the field of computer science as well as the stored-program model of the computer which forms the backbone of the ongoing digital revolution.
In the field of computer science, it led to the systematic study of algorithms and their computational resources.

With the help of a precise automata model, it was now possible to rigorously define the concepts of scientific models using algorithmic information.
Algorithmic information theory (AIT) allows studying the inherent structure of objects without reference to a generating distribution (often assumed erroneously as the uniform prior in statistical machine learning). 
The theory originated when Ray Solomonoff~\cite{solomonoff1964formal}, Andrey Kolmogorov~\cite{kolmogorov1968three}, and Gregory Chaitin~\cite{chaitin1966length} looked at information, probability and statistics through the lens of the computational automata models. 
The theory has now become a central part of theoretical computer science~\cite{li2008introduction}.
While AIT metrics are ubiquitous, most metrics are uncomputable.
Estimating these using approximate methods is often intractable beyond small cases even on classical supercomputers.
Thus, while in the field of theoretical computer science, these fundamental concepts are valuable for proofs, their applicability to real-world data and use-cases remain very limited.

There are various algorithmic metric that will be relevant in the context of this article and will be introduced as and when required.
Ray Solomonoff founded the theory of \textit{universal inductive inference} based on \textit{algorithmic probability} (the chance that a randomly selected program will output a specific output when executed on a TM) and \textit{algorithmic complexity} (the length of the shortest program that outputs a specific output when executed on a TM).
The \textit{universal a priori probability distribution} is the distribution of the algorithmic probability of all strings of a specific size.
This mathematically formalizes the notion of Occam's razor and Epicurus' \textit{principle of multiple explanations} using the computing framework.
When this a prior distribution is used for Bayesian inference, it explains observations of the world by the smallest computer program that outputs those observations, thus, all computable theories which describe previous observations are used to calculate the probability of the next observation, with more weight put on the shorter computable theories.

Although these metric are uncomputable, Levin's \textit{universal search}~\cite{levin1973universal}, converges to the distribution sequence when executed for longer periods of time.
It solves inversion problems by interleaving the execution of all possible programs on a universal Turing machine, sharing computation time equally among them, until one of the executed programs manages to solve the given inversion problem.
This has inspired various artificial general intelligence approaches that built upon this to calculate the expected value of an action.
The more computing power that is available, the closer their predictions are to the predictions of inductive inference.
Jürgen Schmidhuber and Marcus Hutter developed many AGI algorithms like Adaptive Levin Search, Probabilistic Incremental Program Evolution, Self-Modifying Probabilistic Learning Algorithms, Hutter Search, Optimal Ordered Problem Solver, AIXItl and Gödel Machine.

\subsection{Artificial general intelligence}

The last few decades has been a witness to the revolutionary impact of \textit{artificial intelligence} (AI) in many specialized domains.
These diverse domains include medical diagnostics, photo-editing, navigation, to name a few.
AI in this context is an optimization and learning procedure, distinct from the previous generation of explicitly coded algorithms.
These AI algorithms (called, software 2.0) searches for optimal parameters of a context-specific program template that minimizes some defined cost function of the application.
These advances come under the purview of \textit{artificial narrow intelligence} (ANI), where the algorithm is trained to surpass human experts at a specific task (e.g. cancer detection, playing chess) while being completely clueless at a different often simpler task (e.g. logical reasoning, object manipulation).
The impressive success of ANI has given researchers the thrust to strive towards \textit{artificial general intelligence} (AGI), where tasks across domain can be learned, similar to human intelligence.
Currently, state-of-the-art AGI systems can master multiple games~\cite{schrittwieser2020mastering} or converse~\cite{brown2020language} in diverse contexts, without being explicitly programmed with the underlying rules.

Though AGI is closer to the original conceptualization from the early proponents~\cite{nilsson2009quest} of AI (like John McCarthy, Marvin Minisky, John Holland, Ray Solomonoff, and others), we do not have an agreed model of AGI yet.
There are various tribes of researchers, each with a different approach (or, master algorithm~\cite{domingos2015master}) converging towards AGI.
These include symbolists, connectionists, evolutionaries, Bayesians and analogizers.
For example, in \textit{universal artificial general intelligence} (UAGI)~\cite{hutter2004universal}, the primary method is to infer models based on a universal computing model like the Turing machine or lambda calculus.
Hybrid solutions, like neuro-evolution, neural Turing machine, OpenCog, are also being investigated.
The multiplicity of approaches in not because of fundamental incompatibilities preventing an unification (e.g. in theoretical physics), but rather because of the capability of each perspective to converge to AGI given enough resources, making it a matter of design choice.
The model presented in this article follows a hybrid approach based on universal computation, evolutionary programming, compression and probabilistic inference.

The trajectory of ANI and AGI can be extrapolated to artificial super-intelligence (ASI), beyond the point of technological singularity.
ASI has grown out of the domains of scientific fiction to creditable discussions in AI conferences between futurists.
AGI and ASI are together called \textit{Strong AI}, in contrast to Weak AI based on ANI.
In this article, we explore what might eventually one of the mainstream objectives~\cite{hein2018artificial} of Strong AI - making scientific advances that are explainable to human intelligence.
Is it possible to create a general framework for scientific discovery - for both prediction and modeling?
We explore the characteristics of this framework for a feasible experimental setting.

\section{Survey of related methods} \label{s3}

\subsection{General reinforcement learning}

UAGI starts with algorithms or agents that would yield incredibly powerful general intelligence if supplied with massively, unrealistically computing power, and then examines practically feasible (maybe no longer universal) AGI systems as specializations of these powerful theoretic systems.
These algorithms are variations of Solomonoff's universal predictors, which provide a rigorous and elegant solution to the problem of sequence prediction via AIT. 
The core idea is that \textit{the shortest program computing a sequence, provides the best predictor regarding the continuation of the sequence}.

Based on the requirement, the learning process can be put into various settings like supervised, unsupervised, reinforcement, active, lifelong, incremental, lazy, offline, etc.
In the standard \textit{reinforcement learning} (RL) framework, the agent and environment play a turn-based game and interact in steps/turns.
At time step $t$, the agent supplies the environment with an action $a_t$.
The environment then performs some computation and returns a percept $e_t$ to the agent, and the procedure repeats.
There are two main approaches to RL: model-free RL (e.g. Q-Learning, Deep Q-networks) generally assume the environment is a finite-state Markov decision process (MDP) while model-based RL (e.g. Bayesian learning) assumes the realizable case, i.e. \textit{the hypothesis space contains the true function}.
In this research, we follow the framework of Bayesian model-based RL as an interaction between an (artificial) physicist and an environment that it intends to scientifically model and predict.

In fully-observable games (e.g. chess, go), devising an optimal strategy is trivial given an infinite computing power by using the minimax decision rule to enumerate the entire game tree.
In contrast, there is no general-purpose algorithm for the case of decision-making under uncertainty and with only partial knowledge. 
Environments can have traps from which it can be impossible to recover, thus, without knowing a priori which actions lead to traps, it is not trivial for an agent to learning and perform optimally.
This setting is called \textit{general reinforcement learning} (GRL).
There are various theoretical models of artificial general intelligence (AGI) that belong to the class of GRL agents.

\subsubsection{AIXI and its variants}

Some general terminologies are first introduced before presenting the AIXI based agents.
The actions $a_t$ of a GRL agent belong to a (usually finite) set called the action space $\mathcal{A}$.
Likewise, the percepts $e_t$ from the environment belong to the percept space $\mathcal{E}$.
For simplicity, these spaces are assumed to be stationary (i.e. time-independent and fixed by the environment) and countable (although most results generalize to continuous spaces).

The agent is formally identified by its policy $\pi$, which in general is a distribution over actions for the current step, conditioned over the history, denoted by, $\pi(a_t|ae_{<t})$.
The finite sequence of action-percept pairs over the past time steps, $ae_{<t}=a_1e_1\dots a_{t-1}e_{t-1}$ is called the history.
The environment is modeled as a distribution over percepts, $\nu(e_t|ae_{<t}a_t)$.
An environment in GRL is modeled as a partially observable Markov decision process (POMDP).
Thus, there is some underlying (hidden) state space $\mathcal{S}$, with respect to which the environment's dynamics are Markovian.
The agent cannot observe this state directly, but instead receives (incomplete and noisy) percepts through its sensors.
Therefore, the agent must learn and make decisions under uncertainty in order to perform well.

Percepts consist of (observation, reward) pairs,  $e_t=(o_t,r_t)$.
The return is defined as the discounted sum of all future rewards, $R_t = \sum_{i=t}^{m} \gamma_i r_i$.
$m:\mathbb{Z} \in \{t+1,\infty\}$ is the remaining duration the agent is run or upto a certain sliding window of future time steps, called the horizon.
$\gamma:\mathbb{N} \in [0,1]$ is a discount function with convergent sum.
A rational agent based on Von Neumann–Morgenstern utility theorem strives to maximize the expected return, called the value.
The value achieved by a policy in an environment given a history is defined as: $V_\nu^\pi(ae_{<t})=\mathbb{E}_\nu^\pi\big[R_t\big|ae_{<t}\big]$.
This can be expressed recursively, as the Bellman optimality equation, $$V_\nu^\pi(ae_{<t})=\sum_{a_t\in\mathcal{A}}\pi(a_t|ae_{<t}) \sum_{e_t\in\mathcal{E}}\nu(e_t|ae_{<t}a_t) \big[\gamma_t r_t + \gamma_{t+1} V_\nu^\pi(ae_{<t+1})\big]$$

\textbf{AI$\mu$}: The optimal value $V_\mu^* = \max_\pi V_\mu^\pi$ is the highest value achieved by any policy in the true environment $\mu$.
AI$\mu$ is the informed agent with the policy $\pi^{\text{AI}\mu}=\arg\max_\pi V_\mu^\pi$.
This policy is practically infeasible, since the environment $\mu$ is not known a priori in the GRL setting.

\textbf{AI$\xi$}: In the realizable case, the true environment $\mu$ is contained in some countable model class $\mathcal{M}$.
A Bayesian mixture over this class, defined as a convex linear combination of environments, is $\xi(e_t|ae_{<t}) = \sum_{\nu\in\mathcal{M}} w_\nu \nu_{e_t|ae_{<t}}$
The weights $w_\nu = Pr(\nu|ae_{<t})$ specify the agent's posterior belief distribution over $\mathcal{M}$ and lie in the interval $(0,1)$ based on Cromwell's rule.
In a Bayesian agent, these beliefs are updated using the Bayes probability rule.

AI$\xi$ is the Bayes-optimal agent for the policy that maximizes the $\xi$-expected return, $\pi^{\text{AI}\xi}=\arg\max_\pi V_\xi^\pi$.
It is a universal, parameter-free Bayesian agent, whose behavior is completely specified by its model class and choice of prior weights.

\textbf{AIXI}: It uses Solomonoff's universal prior for mixing over the model class of all computable probability measures using the Kolmogorov complexity of the environment, $w_\nu = 2^{-K(\nu)}$.
It is the active generalization of Solomonoff induction, resulting in an optimal but incomputable inductive learner.
The environments are usually modeled as programs on a Universal Turing Machine, $U$, which is typically modeled as a monotone TM with 3 tapes, for input (perception), working and output (action).
Distributing the $\max$ and $\sum$ in the recursive equation gives the canonical expectimax equation as, $$a_t^{AIXI} = \arg \lim_{m\rightarrow \infty} \max_{a_t\in \mathcal{A}}\sum_{e_t\in\mathcal{E}}\dots\max_{a_m\in \mathcal{A}}\sum_{e_m\in\mathcal{E}} \sum_{k=t}^m \gamma_k r_k \sum_{q:U(q;a_{<k})=e_{<k*}} 2^{-l(q)}$$

AIXI asymptotically learns how good its policy, i.e. it is on-policy value convergent, but not asymptotic optimality.
Additionally, bad priors can influence the result to be arbitrarily bad.
Besides, since AIXI is only asymptotically computable, it is not a pragmatic algorithmic solution to GRL, and must be simplified in any implementation.

\textbf{AIXI-tl}: In principle, there are an infinite number programs that can be candidate models of the environment.
Also, while evaluating, the programs can enter infinite loops.
Thus, to circumvent these two issues with AIXI, AIXI-tl limits the length of the programs considered for modeling as well as assigns a timeout for computing the action.
Thus, it is a computable version of AIXI.

\textbf{MC-AIXI$_{\text{(FAC-CTW)}}$}: It approximates AIXI more aggressively than AIXI-tl, by modeling the environment as a Factored Action-Conditional Context Tree Weighting instead of programs and uses Monte Carlo tree search. 

\textbf{MDL agent}: It greedily picks the simplest (Minimum Description Length) probable unfalsified environment in its model class and behaves optimally with respect to that environment until it is falsified.
The policy is thus $\pi_{MDL} = \arg \max_{a\in \mathcal{A}} V_\rho^*$.
The chosen environment is the one with the smallest Kolmogorov complexity, $\rho = \arg \min_{\nu\in \mathcal{M}:w_\nu > 0} K(\nu)$.

\subsubsection{Knowledge Seeking Agents}

One of the central open problems in reinforcement learning is identifying the value in exploration verses exploitation given an incomplete knowledge of the environment.
Knowledge Seeking Agents (KSA)~\cite{orseau2014universal} tries to address the asymptotic optimality issues of AIXI in the UAGI context.
The Bayesian reinforcement learner is generalized to a Bayesian utility agent that has a utility function, $u(e_t|ae_{<t}a_t)$.
For a standard GRL, the utility the agent maximizes is the reward, $u(e_t)=r_t$.
Alternatively, using this formalism, it is possible to define agents that care about objective states of the world, and not about some arbitrary reward signal.
These approaches motivates the agent to explore in environments with sparse reward structure via a principled and general approach, removing the dependence on extrinsic reward.
It collapses the exploration-exploitation trade-off to simply exploration.
KSA assigns the utility as information gain of the model.
The goal of these agents are to entirely explore its world in an optimal way and form a model.
The agent gets reward for reducing the entropy (uncertainty) in its model from the 2 components: uncertainty in the agent's beliefs and environmental noise.

\textbf{Square KSA}: For Square KSA, the utility is $u(e_t|ae_{<t}a_t) = - \xi(e_t|ae_{<t})$


\textbf{Shannon KSA}: In Shannon KSA, the utility is $u(e_t|ae_{<t}a_t) = \log \xi(e_t|ae_{<t})$.
Both Square and Shannon KSA are entropy seeking, which are effective only in deterministic environments.
They fail completely in stochastic environments as they get hooked on the noise and fail to explore.

\textbf{KL KSA}: The Kullback-Leibler KSA's utility function is given by the information gain, as $u(e_t|ae_{<t}a_t) = Ent(w|ae_{<t+1})-Ent(w|ae_{<t}a_t)$.
KL KSA~\cite{orseau2013universal} works even in stochastic noise unlike entropy KSA.

\textbf{BayesExp KSA}: These are knowledge seeking agents with asymptotically optimality unlike AIXI. It augments AIXI with bursts of exploration, using the information seeking policy of KL KSA. If the expected information gain exceeds a threshold, the agent will embark on an information-seeking policy for one effective horizon

\textbf{Thompson sampling KSA}: These are also knowledge seeking agents with asymptotically optimality unlike AIXI. It follows the $\rho$-optimal policy for an effective horizon before re-sampling from the posterior. This commits the agent to a single hypothesis for a significant amount of time, as it samples and tests each hypothesis one at a time.

\textbf{Multi-agent AIXI}: It solves the  grain of truth problem~\cite{leike2016formal} in multi-agent using Thompson sampling. For infinitely repeated games states, as long as each agent assigns positive prior probability to the other agents' policies (a grain of truth) and each agent acts Bayes-optimal, then the agents converge to playing a Nash equilibrium. It was shown to work for both unknown environment and unknown priors of agents.

\textbf{Inq}: Inquisitive Reinforcement Learner~\cite{cohen2019strongly} is another recently proposed UAGI model of strongly asymptotically optimal agent, instead of weakly asymptotic like BayesExp and Thompson sampling. It is more likely to explore the more it expects an exploratory action to reduce its uncertainty about which environment it is in.



\subsubsection{UAI and UCAI}

Unlimited AI~\cite{katayama2018computable} is a further generalization of AIXI and KSA beyond the Turing machine model, for which the rewards need to be rational numbers (AIXI considers real numbers). The computation model need not be TM, but any set of programs taking a lazy list of actions and returning a lazy list of perceptions. The computation is not sequencial like AIXI, but can be parallel.

UCAI is a computable version of UAI which only considers sets of terminating programs, thus, not Turing-complete.

\subsection{Bio-inspired computation}

Biologically inspired computing use models of biology to solve computer science problems.
It uses an evolutionary approach, as opposed to the creationist approach in traditional AI.
Bio-inspired computing is bottom-up and evolves from a set of simple rules and organisms while AI systems are often programmed from above.
We briefly review some important concepts in this discipline that will be required for merging techniques within the QKSA agent. 

\textbf{Evolutionary computation}: It is a family of population-based trial and error problem solving algorithms used as a metaheuristic or stochastic optimization.
They are inspired by biological evolution and are widely used in the fields of artificial intelligence and soft computing.
An initial set of candidate solutions is generated and iteratively updated. 
Each new generation is produced by selecting more desired solutions based on a fitness function, and introducing small random mutations. The population mimics the behaviour or natural selection and gradually evolves to increase in fitness.
Different variants like evolutionary strategies, genetic algorithms, evolutionary programming and genetic programming, were developed to suit specific families of problems and data structures.
There are other metaheuristic optimization algorithms that also are also categorized as evolutionary computation, like agent-based modeling, artificial life, neuro-evolution, swarm intelligence, memetic algorithms, etc.

\subsubsection{Genetic programming}

Alan Turing proposed the idea of evolving program which eventually led to John Holland and John Koza's work on establishing this field of research.
Genetic programming (GP)~\cite{koza1992genetic} is a heuristic search technique of evolving programs, starting from a population of (usually) random programs, for a particular task.
Computer programs in GP are traditionally represented in memory as tree structures (as used in functional programming languages) which can be easily evaluated in a recursive manner. 
The operations applied are analogous to natural genetic processes: selection of the fittest programs for reproduction (crossover) and mutation according to a predefined fitness measure, usually proficiency at the desired task. 
The crossover operation involves swapping random parts of selected pairs (parents) to produce new and different offspring that become part of the new generation of programs. 
Mutation involves substitution of some random part of a program with some other random part of a program. 
Termination of the recursion is when some individual program reaches a predefined proficiency or fitness level.
GP has been successfully used as an automatic programming tool, a machine learning tool and an automatic problem-solving engine, and is especially useful in the domains where the exact form of the solution is not known in advance or an approximate solution is acceptable.

\textbf{Meta-genetic programming}: It was proposed by Jürgen Schmidhuber as a meta learning technique of evolving a genetic programming system using genetic programming itself. 
It suggests that chromosomes, crossover, and mutation were themselves evolved, therefore like their real life counterparts should be allowed to change on their own rather than being determined by a human programmer. 

\textbf{Cyc}: Doug Lenat's Automated Mathematician~\cite{lenat1976artificial}, Eurisko~\cite{lenat1983eurisko} and Cyc are implementation effort with similar scope. 
Cyc is originally an AGI attempt to assemble a comprehensive ontology and human common sense knowledge base that spans the basic concepts and rules about how the world works. 
Though it has been criticized for the amount of data required to produce a viable result and its inability to evolve, Cyc has many successful applications in knowledge engineering and inference.

\subsubsection{Self-replicating programs}

Artificial life (alife) examine systems related to natural life, its processes, and its evolution, through the use of simulations with (soft) computer algorithms, (hard) robotics, and (wet) biochemistry models.
The commonly accepted definition of life does not consider any current alife simulations to be alive, though \textit{strong alife} proponents like John von Neumann advocates that life is a process which can be abstracted away from any particular medium.
The field of soft-alife was mainly developed using cellular automata, while neuro-evolution is another popular technique in use today.
The symbiosis between learning and evolution is central to development of instincts in organisms with higher neurological complexity, e.g. the Baldwin effect.

\textbf{Cellular automata}: Cellular automata (CA) is a discrete model of computation.
A cellular automaton consists of a regular n-dimensional grid of cells, each in one of a finite number of states.
Relative to each cell, a set of cells called its neighborhood is defined.
The computation proceeds according to a fixed rule that determines the new state of each cell in terms of the current state of the cell and the states of the cells in its neighborhood.
Typically, the rule is fixed and same for all the cells and is applied to the whole grid simultaneously.
This have found various applications in physics, chemistry, biology and modeling.
Stanislaw Ulam and John von Neumann~\cite{neumann1966theory} were the originators of this idea and were later popularized by John Conway's Game of Life, and Stephen Wolfram's systematic study~\cite{wolfram2002new} of one-dimensional elementary cellular automata.
Some classes of rules can be shown to be computationally universal.
Different variants of CA relax the criteria and extend it to irregular grid, continuous spaces, probabilistic rules, reversible rules, as well as to quantum rules.

\textbf{Universal constructors}: Universal constructor is a self-replicating machine foundational in automata theory, complex systems and artificial life.
John von Neumann was motivated to study abstract machines which are complex enough such that they could grow or evolve like biological organisms.
Defining this mechanism in more detail was the primary motivation for the invented of the concept of cellular automaton.
The simplest such machine, when executed, should at least replicate itself. 
Additionally, the machine might have an overall operating system (which can be part of the world rule or compiler) and extra functions as payloads.
The payload can be very complex like a learning agent or an instance of an evolving neural network.
The design of a self-replicating machine consists of:
\begin{itemize}[nolistsep,noitemsep]
    \item a program or \emph{description} of itself
    \item a \emph{universal constructor} mechanism that can read any description and construct the machine or description encoded in that description
    \item a \emph{universal copier} machine that can make copies of any description (if this allows mutating the description it is possible to evolve to a higher complexity)
\end{itemize}
The constructor mechanism has two steps: first the universal constructor is used to construct a new machine encoded in the description (thereby interpreting the description as program), then the universal copier is used to create a copy of that description in the new machine (thereby interpreting the description as data).
This is analogous to the cellular processes of DNA translation and DNA replication, respectively.
The cell's dynamics is the operating system which also performs the metabolism as the extra functions when it is not reproducing.

\textbf{Quines}: A quine is a program which takes no input and produces a copy of its own source code as its output.
Thus, it is akin to the software embodiment of constructors.
Quine may not have other useful outputs.
In computability theory, such self-replicating (self-reproducing or self-copying) programs are fixed points of an execution environment, as a function transforming programs into their outputs.
These fixed points are possible in any Turing complete programming language, as a direct consequence of Kleene's recursion theorem.
Quines are also a limiting case of algorithmic randomness as their length is same as their output.
In principle, any program can be written as a quine, where it (a) replicates it source code, (b) executes an orthogonal payload which serves the same purpose the original non-quine version.

Understanding the mechanisms of life as information processing is an early idea~\cite{schrodinger1992life} which eventually led to the discovery of the DNA, and currently~\cite{walker2017matter} used widely for astrobiology and SETI research.
The idea of using the fixed-point, called the Y-combinator in lambda calculus $\lambda f.(\lambda x.f(xx))(\lambda x.f(xx))$ to describe the genetic code~\cite{chaitin2004meta} is pioneered by Gregory Chaitin~\cite{chaitin2012proving} as the field meta-biology.
In the field of transcendental/recreational programming, the DNA structure was used to code the Gödel number (similar to the description number) of any Ruby script~\cite{endoh_2020}.
The space and probability of constructors~\cite{sipper2001go} can inform the subset of DNA encoding for in vitro experimentation and understanding of causal mechanisms in cells~\cite{brenner2012life, condon2018will}.

\textbf{Constructor theory}: Constructor theory\cite{deutsch2013constructor,deutsch2015constructor,marletto2015constructor,marletto2016constructor} is a proposal for a new mode of explanation in fundamental physics, proposed by David Deutsch and Chiara Marletto.
It expresses physical laws exclusively in terms of what physical transformations, or tasks, are possible versus which are impossible, and why. 
A task is impossible if there is a law of physics that forbids its being performed with arbitrarily high accuracy, and possible otherwise.
Such counterfactual reasoning in fundamental physics allows physical laws to be expressed using the lens of information and biology.
Physically constructors can be implemented as a heat engine (a thermodynamic constructor), a catalyst (a chemical constructor) or a computer algorithm (a programmable constructor).
Quantum mechanics and all other physical theories are claimed to be subsidiary theories as special cases.
It is still a relatively new idea and requires further exploration and development.

\subsubsection{Neural networks based models}

Artificial neural networks (ANN) are computing systems inspired by the biological neural networks based on a collection of connected nodes of artificial neurons.
Each node, like the synapses in a biological brain, transmit a signal (a real number) to other connected neurons based on a non-linear function of the sum of its inputs.
The edge between the neurons are weighted, whose value is adjusted as the learning proceeds, and determines the strength of the transmitted signal.
Learning in an ANN typically occurs by training using the difference (error) between the processed output of the network (often a prediction) and a target output. 
According to a learning rule the network adjusts its weighted associations to minimize the error value. 
Successive adjustments cause the network to produce output which is increasingly similar to the target output.

ANNs began as an attempt to exploit to perform tasks that conventional algorithms had little success with. 
Instead of on mimicking the architecture of the human brain, ANN research now focuses mostly on empirical results, using various connection topologies, learning methods and transfer functions, and has developed into a distinct field of its own.
There are also hybrid models extending ANN to the universal and the evolutionary methods.
Neuro-evolution using augmented topologies (NEAT)~\cite{stanley2002evolving} and neural Turing machine (NTM) are two such examples respectively.

Neural networks can exhibit universal computation~\cite{siegelmann1991turing} and thus can be an alternative engine for the QKSA, though it is not our focus due to the inherent difficultly in explainability of ANNs.
There are some neural networks which are inspired by models of physics~\cite{vanchurin2020world}, and this might be a promising future direction for the QKSA.
These networks like Hopfield networks, Boltzman machine, Recurrent neural network and Born machines~\cite{coyle2020born} can be considered as an alternative automata for weighing the hypothesis of the environment by the agent.
Due to our focus on an universal computing model and the invariance theorem, for a large enough diversity of the environment and computing capabilities, it does not matter which automata model is used.

\subsubsection{Projective simulation}

Projection simulation (PS)~\cite{briegel2012projective} is a bio-inspired RL framework which allows the agent, based on previous experience (and variations thereof) to project itself into potential future situations. 
The PS uses a specific memory system, called episodic and compositional memory (ECM) and which provides the platform for simulating future action before real action is taken. 
The ECM can be described as a stochastic network of clips, which constitute the elementary excitations of episodic memory. 
Projective simulation consists of a replay of clips representing previous experience, together with the creation of new clips under certain variational and compositional principles. 
The simulation requires a platform which is detached from direct motor action and on which fictitious action is continuously tested.
Learning takes place by a continuous modification of the network of clips, by: adaptive changes of transition probabilities between existing clips (bayesian updating); creation of new clips in the network via new perceptual input (new clips from new percepts); and creation of new clips from existing ones under certain compositional principles (new clips through composition).

As in the standard theory of reinforcement learning, PS builds entirely on experience (i.e. previously encountered perceptual input together with the actions of the agent). 
The agent can re-excite fragments of previous experience (clips) to simulate future action, before real action is taken. 
Thus in this simulation process, sequences of fictitious memory will be created by a probabilistic excitation process, which are evaluated and screened for specific features, leading to specific action.
The ECM provides a reflection and simulation platform which allows the agent to detach from primary experience and to project itself into conceivable situations.

Thus, the PS model establishing a general framework that connects the embodied agent research with fundamental notions of physics. 
projective simulation as a random walk through the space of clips provides such a general framework, which allows generalizing the model to quantum simulation, thereby connecting the problem of artificial agent design to fundamental concepts in quantum information and computation.
The aim and scope~\cite{melnikov2017active,dunjko2017advances} of PS is very similar to QKSA and will be explored in more depth and contrasted with the QKSA in our future research.


\subsection{Features of a baseline model}

In the last section we presented various GRL and bio-inspied models each with its own advantages and limitations.
Our choice to focus on the UTM as the computational automata eliminates the UAI and UCAI.
While both neural network and projective simulation are alternative approaches to the underlying automata, our current focus is using the AIT metrics on the Turing machine, which is comparatively well developed in this respect.
The multi-agent scenarios are also very interesting for our research, and we leave it for future exploration as extensions.

The KSA based models are a generalization of AIXI where the utility function is defined as some function of the information from the actions and perceptions, instead of only the reward.
We will discuss in the following section the requirement for self-assigned reward for our model, which prevent using the standard AIXI model to be directly applied.
Since quantum environments are inherently stochastic, the entropy based KSA like Shannon-KSA would fail miserably as new entropy can always be extracted from a repeated quantum experiment.
Thus, while though the specifics of our model is considerably different from the ones presented here, in principle it shares more similarity with the KL, BayesExp, Thompson and Inq KSA agents.

The bio-inspired approaches will be used to eliminate some of the limitations of the AIXI and KSA agents in asymptotic optimility.
This method is popularly called the Alife path to AGI.
Genetic programming will be an important feature for evolving the cost function in weighing the candidate hypothesis.
Though we will consider only asexual reproduction with mutation.
Various hypothesis will be active at any moment, thus, a population of QKSA will simultaneously strive to converge on a winning strategy for modeling the environment.
These population of QKSA is produced using a Quine, thus, when a new strategy (cost function) needs to be evaluated on the environment, the QKSA self-replicates with the mutated strategy as part of the offspring.

\section{Quantum artificial intelligence} \label{s4}

Quantum artificial intelligence (QAI) is an umbrella term exploring the synergy between these two disciplines and can be broadly categorized as using (i) quantum physics and technology for AI research and, (ii) using AI for quantum physics and technology research.
In our past work~\cite{sarkar2021estimating}, we have proposed general approaches for the former, where we presented a general quantum circuit based framework that finds application is inferring causal models.
Various other recent research~\cite{benedetti2021variational} focus in this direction.
Due to the huge thrust in classical machine learning, QAI is often wrongly associated with quantum machine learning (QML), the latter being a subset of these techniques.

Any AI setting~\cite{silver2021reward} can be viewed as an interaction between an agent and an environment using reinforcement learning.
Thus, quantum reinforcement learning (QRL) can be structured based on the type of the agent and the environment.
Both the agent and the environment can be enhanced by quantum dynamics or computational abilities, allowing 4 scenarios: $[A_C:E_C]$, $[A_C:E_Q]$, $[A_Q:E_C]$ and $[A_Q:E_Q]$.

\textbf{$[A_C:E_C]$} This includes most of the work described in background sections and serves as a baseline for comparing our model.
Since classical dynamics/computation is a special case of quantum, an agent capable of learning in the other 3 scenarios should be directly translatable.
For UAGI based GRL, simple implementable environment includes games like tic-tac-toe, pac-man, grid navigation, etc.
Classical physics can also be learnt in a RL setting and is the basis for mechanical models used in animations and robotics.

\textbf{$[A_C:E_Q]$} A quantum environment can be learnt by a classical agent via measurements.
This is possible due to the Church-Turing thesis allowing modeling any physical process (albeit with an exponential cost) as quantum processes are not super-Turing.
This will be the main focus of this article and will be discussed in more detail later.

\textbf{$[A_Q:E_C]$} This includes employing quantum computations to accelerate the inference procedure for the agent.
AIXI-q~\cite{catt2020gentle} is a proposed quantum upgrade for the AIXI model, though it has not been implemented yet and there are some details that needs to worked out before it could be models in current quantum systems.

\textbf{$[A_Q:E_Q]$} This is the most general setting where both the environment and agent is quantum.
It includes concepts from quantum information theory, which bounds the amount of shared information and includes paradoxes like the Wigner's friend scenario.
These will be explored in a future work as an extension where our model can be applied.

\subsection{Quantum machine learning}

Quantum machine learning has been a very promising direction of research in recent year.
Broadly, it can be distinguished into two approaches.

\textbf{QC for ML}: In the field of quantum algorithms, QML refers to quantum solutions of machine learning problems, including neural networks, clustering, regression, optimization, etc.
They seek to find algorithms that provides a benefit in terms of runtime, trainibility, solution quality, or memory space, with respect to existing solutions.
This is beyond the scope of the research presented in this article.

\textbf{ML for QC}: Machine learning techniques has been used to optimize and control processes in the development of quantum computers.
These include control of the quantum system, routing and mapping of qubits, quantum error-correction~\cite{varsamopoulos2019comparing}, etc.
Most of these techniques are based on neural network.
Restricted Boltzmann machines (RBM) has also been used for learning of quantum states~\cite{torlai2018neural} and processes, and is thus very similar to the application of QKSA discussed in this article.
Melvin~\cite{krenn2016automated} is a computer algorithm which is able to find new experimental implementations for the creation and manipulation of complex quantum states. 
The classical optimizer in quantum variational approaches has also been implemented as a reinforcement learner~\cite{wauters2020reinforcement,yao2020reinforcement,rivera2021avoiding,carrasquilla2021neural}.
However, these approaches are not general to meet the definition of an AGI agent in a RL setting, and were specifically designed for the application.
Projective simulation is an alternate approach with similar aims as the QKSA, though not based on the universal computing model.

\subsection{Classical observers in a quantum world}

A primary objective of the QKSA is to mimic the human behavior to form explainable hypothesis about the environment.
An inherent feature of semantic explanation in terms of human knowledge is to restrict the representation of knowledge in terms of classical information.
Note that this does not restrict representing quantum information, as using the standard formalism we can represent quantum information using a worst-case exponential amount of real-valued classical information.
Thus, more specifically via the QKSA, we intend to model classical observers (like humans) in a quantum world, and still recover and learn features that can help us form hypothesis and predict the environmental dynamics.

\subsubsection{It from (qu)bit}

\textit{Digital physics} is a hypothesis that the universe can be conceived of as a vast, digital computation device, or as the output of a deterministic or probabilistic computer program.
This program for a universal computer that computes the evolution of the universe could be, for example, a huge cellular automaton.
It was advocated by Konrad Zuse~\cite{zuse1970calculating}, Edward Fredkin, Tommaso Toffoli and others.

John Archibald Wheeler symbolized this idea as \textit{It from bit}.
This meant that every item of the physical world has at bottom an immaterial source and explanations of what we call reality arises from the posing of yes-no questions and the registering of equipment-evoked responses.
It advocates that all physical things are information-theoretic~\cite{wheeler2018information} in origin and this is a participatory universe where the interactions define the reality we perceive.
Based on our modern understanding of quantum information theory as a generalization of Boolean logic, this idea is popularized as \textit{It from qubit} by Seth Lloyd~\cite{lloyd2006programming}.

Information is increasingly put into the central stage in physics, especially in reconstructing theories like quantum mechanics from general principles~\cite{hardy2001quantum, masanes2011derivation,hohn2017quantum} as well as it's physical nature~\cite{vopson2019mass}.

\subsubsection{Law without law}

Wheeler asked the question of the possibility of the existence of an ultimate law of physics, from which everything that is knowable about the material world can be deduced.
This idea has been coined as \textit{law without law}~\cite{deutsch2017wheeler}. 
If such a principle does not exist, it would mean that there exist aspects of the natural world that are fundamentally inaccessible to science.
Instead, if such an unified law exist, then the problem of its own
origin comes into question, like why that particular principle rather than some
other
So, paradoxically, the ultimate principle of physics, cannot be a ``law" (of physics), hence the expression.

Pursuing this idea eventually leads to epistemological assumptions of how physical theories are formed and verified.
This is now increasingly being incorporated into physics, removing physics as the science of ``what is" to that of ``what we observe"

\section{Quantum Knowledge Seeking Agent} \label{s5}

The recent work from Markus Müller~\cite{mueller2020law} is central to the ideas developed in this research.
As a matter of fact, this research is an implementation and extension of the ideas presented in Müller work based on the ``Law without law" idea of John Wheeler.
We briefly review the core idea explored in his work.
In this research, Mueller claims that given a complete description of the current observer state $x$, it is possible to predict what state $y$ the observer will subsequently evolve to using $P(y|x)$ based on Solomonoff's algorithmic probability, universal prior and universal induction.
This currently encompasses classical (non-relativistic) and quantum physics, and can be used to reconstruct an operational theory based on this assumption of the world being computable, and there are no super-Turing physical processes.
These epistemic derivation of the laws of physics based on information axioms focus on using the theories to predict the results of future interactions, instead of the ontological interpretations of the model used to predict.

In this research we extend the research from Markus Müller~\cite{mueller2020law} towards an implementation framework.
While doing so, we narrow down on the specifics of the original ideas.

\subsection{QPT as a general modeling technique}

Extending Müller's idea to the Church-Turing-Deutsch thesis, the program (that the Solomonoff induction uses) is basically an efficient quantum computing simulator, given a classical computing substrate, or a programmable quantum simulator (as Feynman originally imagined) given a quantum computing substrate.
This allows creating a model of the environment (universe) from the agent's (observer's) perspective.
Without going into ontological debates, we maintain that the model is representative of the black-box input-output behaviour of the environment.

Given knowledge of the environmental dynamics, it is possible to create the corresponding classical model (e.g. a Grover search simulator). 
However, for unknown environment, the general technique is to do process tomography, thus, that is the general modeling algorithm we would focus on.

Thus, for the quantum case, we intend to understand what kind of algorithms would execute for predicting the next observer state using Solomonoff universal induction. 
Given that it is possible to simulate quantum physics on a classical simulator (albeit by incurring exponential resource cost), the predictor will be some form of a quantum process tomographic reconstruction based on the previous observer states. 
Thus, an AGI agent trying to derive the information based operational laws of quantum physics would converge to a QPT algorithm based on a pareto-curve of resource constraints defined by the \textit{least} metric (discussed later).
In this research, we propose a meta-learning platform to evolve the AGI agent using GP/Quines and GRL.

\subsection{Participatory observer as a RL agent}

Consider the phase before the process matrix has been reconstructed upto a certain degree of precision (i.e. we have an informationally complete history of observations). 
In this phase, the participatory observer can choose an action like an AIXI agent based on the process matrix, and the observation will be based on both the chosen action and the environment. 
Thus, it is not fully modeled by Solomonoff's induction. 
In the reinforcement learning setting, the next observer state is based on both the current observer state (that defines the memory of previous observations and the current action based on the QPT scheme) as well as the part of the environmental dynamics that has not yet being learnt. 

Note, given a complete description of the environmental dynamics already learnt and encoded within the observer state $x$, it is still possible for the agent to choose (stochastically/free-will) a measurement basis that would reflect in the observation. 
This is called the ``Participatory observer" RBQ. 
It is not a violation of Müller's hypothesis as the scheme (stochastic or deterministic) is still dictated by the previous observer state (with an additional access to a random tape in the TM) and can be encapsulated as part of the QPT algorithm. 

\subsection{Computational resource bounds}

Considering computational resources, to predict the environment, we need to make a model of the environment, that would use comparable amount of resources (e.g. for 2 qubit environment, we would need $2^2$ complex numbers, or 2 qubits, within the agent). 
Often this cannot be possible (e.g. modeling the universe in cosmology), and thus there needs to be a metric that trades-off bounds on resources (like space, time, approximations) to create a pareto-optimal frontier of models and algorithms that can be used to predict. 
There is no good way to create these trade-off function other than a RL or GP function by evaluating their fitness in prediction.
What is also interesting to investigate is if there is a fundamental invariant cost function (similar to area-laws in holography) on these resources which physical modeling trade-offs by reducing the description length and increasing the other parameters.

Not all quantum process has a short classical description, e.g. the description length of a rotation gate $RX(\theta)$ depends on the number of significant bits in the specification of $\theta$.
Given a limit on the length of the classical description (e.g. QASM file length), the complexity of the quantum state cannot be arbitrary.
Thus, given the complexity of the quantum state is not infinite, we can converge to an estimate with bounded error with a finite number of tomographic trials, and can be linked to the formalism of \textit{quantum Kolmogorov complexity}.
The trials need not be equally distributed over the tomographic basis, as the description complexity might be also biases towards a specific basis, e.g. for a state on the XY plane on the Bloch sphere, there is no use measuring in the X basis as it is always random.
In this research we explore a strategy for the agent to learn these insights.

AIXI has inspired other computable variants like AIXI-tl, MC-AIXI$_{\text{(FAC-CTW)}}$ and UCAI.
These models consider bounds on the program length $l$ and the program run time $t$.
Note that while these do not explicitly bound the working memory/tape, the causal cone of the tape gets bounded to $s=2t$ due to the run time bound of $t$.
In retrospect, this is the reason the complexity class of PSPACE is much larger than PTIME.
In many algorithms, it is a common practice to use space-time trade-offs, where more memory can be used to reduce the execution time, or vice versa.
Two additional metrics are considered in our model: the approximation $a$ and the energy consumption $e$, both of which are becoming very important for current computing systems.
The approximation directly links to the reward function, as it sets a discount on the distance penalty between the prediction and the perception.
Estimating the energy is much more difficult to do, as it involves the thermodynamics cost of Turing machines.
In principle, it is done either by counting the number of elementary operations or by directly measuring the unit of power consumed while running the algorithm on a dedicated system.
Some recent work are looking into the mathematical model of estimating this cost from AIT metric.

These metrics together is called the \emph{least} metric, as an acronym for (program/hypothesis) length, (compute) energy, approximation, (work memory) space and (run) time.
The estimates of the \emph{least} metric for the hypothesis is used in a two-fold way.
Firstly, it is used to qualify the hypothesis for consideration based on an upper bounds (set as evolving hyper-parameters) for each of the five metric individually.
Then, the metric for valid hypothesis are fed to a cost function (a genetic program) that outputs a single positive real value which is used as the weight for the hypothesis in the semi-measure instead of only the length, as in algorithmic probability.

\subsection{Multi-agent interpretation of QM}

Quantum mechanics is often criticized as a single-user theory, which does not tell what will be the consensus of an observation for multiple agents.
However, this should not be viewed as a limitation as other successful theories of physics like general relativity also advocate observables as an interaction between the observer and the environment.
Modern epistemic interpretations of QM, like relational quantum mechanics (RQM) and QBism strives to formalize this as part of the description.

Given that we use QPT as the basis for modeling the environment, it gives us a general framework to discuss multi-agent interactions like Wigner's friend. 
The AGI agent (with all the complexities of the QPT and Quines) still has an equivalent process matrix, and thus another agent can in principle model both the environment and the 1st agent. 
Thus, this can be a testing ground from QBism and RQM. 
Similar extensions can also be constructed for causal indefiniteness.

\subsection{Reward function for modeling}

A reinforcement learning model is typically defined as learning the association between a set of action and perception as interactions between an agent and an environment.
To guide the learning process, the model requires a reward function.
Typically, for example in models like AIXI, the environment assigns a reward for the action of the agent at each step.
However, this requires the environment to be aware of the objective function as well as the experimental setup of the agent.
While this is trivial for game environments, it becomes difficult to define for modeling environment dynamics.


To circumvent this constraint, we define the reward as a value that is computed by the agent, instead of being an external input.
This is a generalisation of the AIXI model, since, the reward function that the agent uses to compute can be only a function of the perception from the environment.
In this work we will focus on model-based RL, which would provide a prediction from the environment model corresponding to the same action that the agent performs on the real environment.
The reward is estimated internally by the agent based on a self-defined distance function based on the perception from the environment and the prediction from the model.
This value of the reward is subsequently used by the agent to update the environment model, the distance function itself, as well as inform the next action that would steer the learning objective.
The typically learning objective in RL is to maximize future rewards, which in this case corresponds to converging on a model of the environment with a action-perception association similar to the real environment.
The important change is that the environment need not have knowledge of the reward function and can be changed by the agent at any point.

As an example, for modeling physical dynamics of a mechanical system, the agent uses a trial hypothesis based on the laws of motion.
The environment, e.g. a rock rolling down the hill, is blissfully unaware of the hypothesis the agent (say, a physicist) is using.
The agent must itself decide the correctness of the hypothesis, by assessing the accuracy of prediction the hypothesis results.

Note that this is a considerable point of divergence from the AIXI and PS model, as while we can use the trial hypothesis to generate prediction for future time steps, it is not possible to estimate the reward for future time steps without the actual perception.
Therefore, the fitness of the hypothesis is entirely based on past events.

\subsection{Open-ended evolution of the cost manifold}

The least metric assigns 5 bounds, which creates a 5-dimensional hypothesis space for searching for an efficient QPT reconstruction technique.
But how to compare candidate hypothesis that are within this space?
This is answered by searching for a function that defines a pareto-optimal manifold in this space.

This manifold is part of a gene that evolves in the offspring with a mutation rate as a hyper-parameter.
We define a genetic programming setup with these 5 bound as hyper-parameters, 5 estimates of the least metric for the candidate hypothesis, 5 weights for scaling these estimates, and a function set of addition, subtraction, multiplication, division, logarithm and exponent.
The initial seed AGI starts with a simple addition of the 5 estimates with all weights set to $1$.

The self-replication policy is based on the predictive power of the current scheme for QPT.
Given the set of past measurement, if the QPT fails to effectively predict the future experiments, we adjust the QPT algorithm (e.g. adding more memory or precision) in the offspring created by the Quine framework.
If the prediction fails beyond a certain bound, the current QPT scheme is retired.





\section{Formal definition of QKSA} \label{s6}

In this section we present the formal framework of the QKSA by defining the parameters within an implementation that captures the agent scheme discussed so far.
Thereafter, an execution procedure is outlined as a pseudo-code which forms the baseline for the implementation.
The formalization in this section is independent of QPT, which will be discussed in the following section.


\subsection{Parameters definition}

\subsubsection{Hypothesis fitness based on the least metric}
The least metric defines the bounds on the hypothesis-space and the relative weight assigned to each considered hypothesis.
It takes into account the 5 cost metrics of program length, thermodynamic cost, approximation, space/memory and run-time.

\begin{itemize}[noitemsep,nolistsep]
    \item $d$ refers to the data on which the least metric is evaluated. It consists of a sequence of actions and perceptions for the past time steps $t_p$ (defines as part of the learning model), a chosen trial action for the current time step, and a prediction of the perception from the environment for the current time step.
    \item $h$ is the program, hypothesis or model that upon execution on a defined computational model $C$ generates $d$ as the output.
\end{itemize}

The hypothesis-space is bounded by the 5 $\text{least}_{max}$ hyper-parameters.
All trial hypothesis must lie within the bounds of all 5 parameters.

\begin{itemize}[noitemsep,nolistsep]
    \item $l_{max}$ is the maximum length of $h$ that is considered, that outputs $d$.
    \item $e_{max}$ is the maximum energy cost of executing the $h$ to generate $d$ as output.
    \item $a_{max}$ is the maximum approximation threshold that the prediction is allowed to deviate from the perception based on the distance function used for calculating the reward.
    \item $s_{max}$ is the maximum space or working memory that $h$ can use while execution.
    \item $t_{max}$ is the maximum execution time for $h$ before it generates the output $d$ and halts.
\end{itemize}

Once a trail hypothesis $h$ is admitted based on the $\text{least}_{max}$ bounds, an estimate of the 5 cost parameters is computed.
A specific implementation of our model defines an estimation technique for these parameters.

\begin{itemize}[noitemsep,nolistsep]
    \item $l_{est}$ is an estimate of the length of $h$ that is considered, that outputs $d$. A general scalable technique is to used BDM~\cite{soler2014calculating} as proposed by Hector Zenil.
    \item $e_{est}$ is an estimate of the energy cost of executing $h$ to generate $d$ as output. Research into this aspect is scarce. The recent proposal from David Wolpert on the thermodynamic Kolmogorov complexity~\cite{kolchinsky2020thermodynamic} needs to be explored further for estimating the energy cost. 
    \item $a_{est}$ is an estimate of the deviation between the prediction and the perception based on the distance function used for calculating the reward.
    \item $s_{est}$ is an estimate of the space or working memory that $h$ uses while execution.
    \item $t_{est}$ is an estimate of the execution time for $h$ before it generates the output $d$ and halts. Space-time trade-offs like the pebbling game~\cite{meuli2019reversible} proposed by Charles Bennett can be used in forming the optimal manifold. AIT metrics like logical depth, Levin complexity and speed prior can be used to trade-off between the program length and the execution time.
\end{itemize}

The 5 cost parameters $\text{least}_{est}$ now needs to be combined to form a single indicative metric of the fitness of the hypothesis $h$.
Each parameter has an associated weight or scaling factor $w_{\text{least}}$.
The cost function defines the equation to combine the $\text{least}_{est}$ and $w_{\text{least}}$, and is subject to evolution.

\begin{itemize}[noitemsep,nolistsep]
    \item $w=\{w_l,w_e,w_a,w_s,w_t\}$ is a set of associated weights for each of the least metric.
    \item $c$ is the cost function that takes in the 5 least metrics and a weight for each metric.
    \item $c_{est}$ is the estimated cost based on the estimated least metrics and the cost function.
\end{itemize}

\subsubsection{Policy for reinforcement learning model}

The parameters of the reinforcement learning model defines the policy for the action at each cycle.

\begin{itemize}[noitemsep,nolistsep]
    \item $t_p$ is the number of time steps in the past that is considered by the agent at each point in time. It is a function of the total memory of the computational model $s_{c}$.
    \item $t_f$ is the number of time steps that the agent predicts in the future.
    \item $a_t \in \mathcal{A}$ is the chosen action from the action space at time step $t$.
    \item $e_t \in \mathcal{E}$ is the perception recorded by the agent at time step $t$ from the percept space.
    \item $\rho_t \in \mathcal{E}$ is the prediction of the perception $e_t$ made at time step $t-1$.
    \item $\pi_t$ is a policy for choosing an optimal action and maximizing return at time step $t$. For our agent, we consider all possible actions and prediction for the current step, and choose the hypothesis which meets the $least_{max}$ for each metric individually, and has the least $c_{est}$.
\end{itemize}

\subsubsection{Evolution of cost function based on reward}

The parameters for the quine defines when and how the agent self-replicates.
Replication is triggered based on the fitness of the hypothesis based on the predictive capacity over time.

\begin{itemize}[noitemsep,nolistsep]
    \item $m_c$ is the mutation rate of the cost function $c$.
    \item $F$ is the set of functions allowed in the cost function $c$.
    \item $\Delta$ is a distance measure (e.g. Hamming distance) defined between elements in the percept space.
    \item $r_t=-\Delta(e_t,\rho_t)$ is the reward at time step $t$.
    \item $\gamma_t$ is the reward discount that is proportional to the time span $t$ between the reward step and the current time step.
    \item $R_t=-\sum_{i=t-t_p}^{t-1} \gamma_i \Delta(e_i,\rho_i)$ is the cumulative discounted return at time step $t$.
    \item $R_D$ is the return threshold for death. If $R_t < R_D$ the agent halts (dies).
    \item $R_R$ is the return threshold for reproduction. If $R_D < R_t < R_R$, the agent self-replicates with mutation in its hyper-parameters.
\end{itemize}

\subsubsection{Computation models}

Executing the hypothesis $h$ to generate the data $d$ requires a computation model.
Models like AIXI are based on the monotone Turing machine.
The resource-bounded variant for our model limits both time and space resource, thereby, fitting the definition of a universal linear-bounded automata (ULBA).
The hypothesis $h$ in the ULBA context is the finite state machine that generates the desired output and halts.

\begin{itemize}[noitemsep,nolistsep]
    \item $n$ is the alphabet size of the ULBA that the agent uses for modeling.
    \item $m$ is the state size of the ULBA.
    \item $s_{c}$ is the tape size of the ULBA.
\end{itemize}

Instead of the Turing machine model, one can envision a neural network based model using auto-encoders where the latent space defines the hypothesis, and the total number of neurons defines the space cost of the computational model.
Similarly, projective simulation based model can also be beneficial for certain applications.

In the implementation of QKSA, we will create the hypothesis as computer programs for a programming language (e.g. Python).
Thus, the state and alphabet size is not directly discernible.
Since they are not directly related to any other parameters, that is not a concern.
The tape size (required for the bounds on $t_p$) is directly proportional to the total amount of system memory that is allocated to the program.
This can be managed by a quine hypervisor as an alife simulator, that would arbitrate the allocation of memories among the population of agents at any point of time.

\subsection{Execution procedure}

\begin{itemize}[noitemsep,nolistsep]

    \item[] environment-agent interactions
    \begin{enumerate}[noitemsep,nolistsep]
        \item[] define action: \textit{act}(\texttt{a\_t})
        \item[] initialize action space: \texttt{A} 
        \item[] define perception: \texttt{e\_t} $=$ \textit{perceive}()
        \item[] initialize percept space: \texttt{E}  
        \item[] initialize past and future horizons: \texttt{t\_p, t\_f}
    \end{enumerate}
    
    \item[] least model 
    \begin{enumerate}[noitemsep,nolistsep]
        \item[] define estimators: \texttt{l\_est, e\_est, a\_est, s\_est, t\_est}
        \item[] initialize metric weights: \texttt{wt\_l, wt\_e, wt\_a, wt\_s, wt\_t}
        \item[] initialize metric bounds: \texttt{l\_max, e\_max, a\_max, s\_max, t\_max}
        \item[] initialize cost function set and gene: \texttt{F, c} 
    \end{enumerate}
    
    \item[] quine model 
    \begin{enumerate}[noitemsep,nolistsep]
        \item[] define distance metric: \texttt{r\_t} $=$ \textit{Delta}(\texttt{e\_i, e\_j})
        \item[] initialize reward discount factor: \texttt{gamma} 
        \item[] initialize return bounds for death and replication: \texttt{R\_D, R\_R}
        \item[] initialize lifespan: \texttt{lifespan}
        \item[] initialize gene mutation rate: \texttt{m\_g}
    \end{enumerate}
    
    
    \item[] set: \texttt{t}$=0$, \texttt{R\_t} $=$ \texttt{R\_R}
    
    \item[] \textit{while} (\texttt{t} $<$ \texttt{lifespan}):
    \begin{itemize}[noitemsep,nolistsep]
        \item[] \textit{if} (\texttt{R\_t} $<$ \texttt{R\_D}):
        \begin{itemize}[noitemsep,nolistsep]
            \item[] \textit{break}
        \end{itemize}
        \item[] \texttt{c\_t\_star} $= \infty$
        \item[] \textit{def futureCone}(\texttt{t\_fc, past\_a, past\_e}):
        \begin{itemize}[noitemsep,nolistsep]
            \item[] if (\texttt{t\_fc} $<$ \texttt{t}$+$\texttt{t\_f}):
            \begin{itemize}[noitemsep,nolistsep]
                \item[] \textit{for} \texttt{a} \textit{in} \texttt{A}:
                \begin{enumerate}[noitemsep,nolistsep]
                    \item[] \textit{if} (\texttt{t\_fc} $==$ \texttt{t}):
                        \begin{enumerate}[noitemsep,nolistsep]
                            \item[] \texttt{a\_t} $=$ \texttt{a} 
                        \end{enumerate}
                    \item[] \texttt{past\_a\_new} $=$ \textit{deepcopy}(\texttt{past\_a})
					\item[] \texttt{past\_a\_new}.\textit{append}(\texttt{a})
                    \item[] \textit{for} \texttt{rho} \textit{in} \texttt{E}:
                    \begin{enumerate}[noitemsep,nolistsep]
                        \item[] \texttt{past\_rho\_new} $=$ \textit{deepcopy}(\texttt{past\_rho})
    					\item[] \texttt{past\_rho\_new}.\textit{append}(\texttt{rho})
                    \item[] \textit{futureCone}(\texttt{t\_fc}$+1$,\texttt{ past\_a\_new, past\_rho\_new})
                    \end{enumerate}
                \end{enumerate}
            \end{itemize}
            \item[] else:
            \begin{itemize}[noitemsep,nolistsep]
                \item[] \texttt{cost} $=$ \textit{c\_est}(\texttt{past\_a, past\_e})
                \item[] \textit{if} (\texttt{cost} $> 0$) \textit{and} (\texttt{cost} $<$ \texttt{c\_est\_star}):
                \begin{enumerate}[noitemsep,nolistsep]
                    \item[] \texttt{c\_est\_star} $=$ \texttt{cost}
                    \item[] \texttt{a\_t\_star} $=$ \texttt{a\_t}
                \end{enumerate}
            \end{itemize}
                    
        \end{itemize} 
        
        \item[] \textit{futureCone}(\texttt{t, hist\_a, hist\_e})
        
        \item[] \textit{act}(\texttt{a\_t\_star})
        \item[] \texttt{hist\_a}.\textit{append}(\texttt{a\_t\_star})
        
        \item[] \texttt{rho\_t\_star} $=$ \textit{predict}(\texttt{a\_t\_star})
        
        \item[] \texttt{e\_t} $=$ \textit{perceive}()
        \item[] \texttt{hist\_e}.\textit{append}(\texttt{e\_t})
        
        \item[] \texttt{r\_t} $=$ \textit{Delta}(\texttt{rho\_t\_star,e\_t})
        \item[] \texttt{hist\_r}.\textit{append}(\texttt{r\_t})
        
        \item[] \texttt{R\_t} $= 0$ 
        \item[] \textit{for} \texttt{i} \textit{in range}($0$,\texttt{ t\_p}):
        \begin{itemize}[noitemsep,nolistsep]
            \item[] \texttt{R\_t} $+=$ \texttt{hist\_r[-i-1]} $*$ ($1 -$ \texttt{gamma}$*$\texttt{i})
        \end{itemize}
        
        \item[] \textit{if} (\texttt{R\_t} $<$ \texttt{R\_R}):
        \begin{itemize}[noitemsep,nolistsep]
            \item[] mutate hyper-parameters in gene
            \item[] self-replicate with mutated gene
        \end{itemize}
        
        \item[] \texttt{t} $+= 1$
    \end{itemize}
\end{itemize}

\vspace{1em}
\noindent \textit{c\_est}(\texttt{past\_a, past\_e}) and \textit{predict}(\texttt{a\_t\_star}) embeds the logic of the environment that the QKSA learns.

\section{Quantum process tomography} \label{s7}

In this section we will introduce quantum process tomography (QPT) as a general modeling technique applicable for any classical and quantum environment given enough computation power.
The task of reconstructing quantum states or process from experimental data is variously called quantum state or quantum process tomography, estimation, or recovery.
Since classical universal logic (e.g. NAND gate) can be implemented using quantum logic (e.g. Toffoli gate), quantum processes can efficiently implement classical processes as well.
Note that QKSA does not try to devise ``the most efficient" QPT scheme.
The agent automates the process to intelligently device ``a workable scheme" to estimate the process by choosing preparation and measurement basis for each trail with computational resource assumptions. 
Mutations in the relative weights of the various resources in the least metric generates a population of QKSA each evaluating QPT schemes using a different cost function.

While in principle we want the QKSA to evolve from zero-knowledge, starting with a simple QPT strategy (often called the Seed AI in AGI), is more practical given the current limitations of the field.
Thus, the Seed QKSA is a QPT reconstruction algorithm. 
There are various QPT techniques like standard quantum process tomography (SQPT)~\cite{chuang1997prescription}, ancilla-assisted process tomography (AAPT), entanglement-assisted process tomography (EAPT), direct characterization of quantum dynamics (DCQD), compressed-sensing quantum process tomography (CQPT), permutation-invariant tomography, self-guided quantum process tomography (SGQPT)~\cite{hou2020experimental}, etc.
In the rest of this section we will introduce the density and process matrix formalism which will form the internal representation (IR) for the QKSA of the information learnt from the environment.

\subsection{Quantum states and unitary operators}

Quantum information science is the theoretical framework for computations performed on quantum computers. 
The basic unit of a quantum computer is the quantum bit or qubit, which unlike classical bits are not restricted to either 0 or 1, but can also be in a quantum mechanical superposition: a linear combination of the two states.
Note that this can be generalized to qudits as d-level systems.
In quantum mechanics, particles or systems can be in a non-classical combination of classical states, called the superposition principle, which can be viewed mathematically as a vector in $\mathbb{C}^2$ (called the Hilbert space $\mathcal{H}$) often written using the ket-notation as $\ket{\psi} = \alpha_0 \ket{0} + \alpha_1 \ket{1}$, where $\alpha_i \in \mathbb{C}$.
$\ket{i} \in \{\ket{0},\ket{1}\}$ form a set of basis for $\mathbb{C}^2$.
This basis is called the computational or canonical basis.
Upon measurement of the qubit in a given basis, the state of the qubit collapses to either of the basis state with the probability given by the Born rule as $\text{Pr}(\ket{i}) = |\alpha_i|^2$.
Since probabilities add up to 1, the state is normalized as $\sum |\alpha_i|^2 = 1$.

Although the computational basis encompasses the two physically distinct levels of the qubit, the state of the qubit can be expressed using any two linearly independent vectors that form a basis for $\mathcal{H}$, which are normally chosen as two orthogonal vectors (i.e. $\braket{i}{j} = 0$ and $\braket{i}{i} = 1$).
Two other common basis in use is the Hadamard basis (X-basis) and the Y-basis, expressed in terms of the computational basis (Z-basis) as: $\{\ket{+} = \frac{1}{\sqrt{2}}(\ket{0}+\ket{1}), \ket{-} = \frac{1}{\sqrt{2}}(\ket{0}-\ket{1})\}$ and $\{\ket{+i} = \frac{1}{\sqrt{2}}(\ket{0}+i\ket{1}), \ket{-i} = \frac{1}{\sqrt{2}}(\ket{0}-i\ket{1})\}$  respectively. 

The above notation can be generalized to multiple qubits using the Kronecker tensor product on the Hilbert space.
Total state of $n$ qubits is a vector in a Hilbert space of dimension $2^n$ (for qudits, $d^n$).
For example, the $\{\ket{00},\ket{01},\ket{10},\ket{11}\}$ form a basis for 2 qubits.
Thus the number of parameters needed to describe the state grows exponentially with the number of qubits, often quoted as the primary tool in quantum computation to achieve superior computational capability over classical systems.

Quantum computations are performed by means of operations on the n-qubits as a map $\mathcal{O} : \mathcal{H}_n \rightarrow \mathcal{H}_n$ mapping qubit states to other qubit states.
Normally, these operators are unitary, i.e. $U^\dagger U=I$.
These $U$ can be viewed as a change in basis for the state, with $U^\dagger$ as the reverse transform.
The combined operation on two subsystem is the operation $U_{12} = U_1 \otimes U_2$.
If this operation cannot be written as a tensor product of local operations on the subsystems it is called an entangling operation.
Since the gate-set $\{CNOT, H, T\}$ is universal, all unitaries can be decomposed~\cite{krol2021efficient} to single-qubit gates and 2-qubit entangling operations.
Since the $SWAP$ gate can be formed from $CNOT$, all unitaries can in fact be decomposed to 2-qubit local interactions.
Note that these decomposition can incur worst-case exponential cost in number of gates.

\subsection{Density matrix for quantum states}

A density matrix is a generalization of the more usual state vectors or wave-functions formalism used for describing the quantum state of a physical system.
It can be used for pure as well as mixed states for the calculation of the probabilities of the outcomes of any measurement performed upon this system, using the Born rule. 
Mixed states arise when the preparation of the system is not fully known, or when describing a physical sub-system which is entangled with another.
Since we can only reconstruct the density matrix from quantum state tomography, and not the state, in this research we will use the density matrix formalism to represent quantum states.
Density matrix is defined as a statistical ensemble over a set $\{\ket{\psi}_k\}$ of $N$ pure quantum states.
$$\rho = \sum_{k=1}^{N} p_k \ket{\psi}\bra{\psi}$$
where, $0 < p_k \le 1$ and $\sum_{k=1}^N p_k = 1$


A global phase of a quantum state is undetectable. 
Thus, $\ket{\psi} = e^{i\theta} \ket{\psi}$. But, a density matrix is unique, as the corresponding density matrix of a quantum state, $\rho = \ket{\psi}\bra{\psi}$ cancels out the global phase.
It is formed from all epistemic information measured via observations of the quantum state.
If an unitary $U$ acts on the state $\ket{\psi} \rightarrow U\ket{\psi}$, the corresponding transformation in the density matrix $\rho$ of the state is $\rho \rightarrow U\rho U^\dagger$.
A projective measurement of an observable $M$ on a qubit on a pure state gives an expectation value $\langle M \rangle = \bra{\psi} M \ket{\psi} = Tr(M \bra{\psi}\ket{\psi}) = Tr(M \rho)$

The generalization of projective measurements is called a positive operator valued measure (POVM).
In it the qubit state is not projected upon a set of orthogonal states but instead on a set of orthogonal subspaces that together span the whole Hilbert space. 
For a set of operators $\{A_i\}$, the qubit state will collapse with probability $Tr(A_i \rho A_i^\dagger)$ to any of the states $\rho_{A_i} = \dfrac{A_i \rho A_i^\dagger}{Tr(A_i \rho A_i^\dagger)}$.
$\sum_i A_i^\dagger A_i = \mathbb{I}$, so that if the measurement outcome is not recorded, the mixed state $\sum_i A_i \rho A_i^\dagger$ is obtained.
In this work we will deal only with projective measurements.

\subsection{Choi matrix for quantum processes}

A quantum process $\mathcal{E}$ that transforms a density matrix need not be always unitary.
Given classical processes are often irreversible and includes measurements, a quantum generalization should include unitary transforms (symmetry transformations of isolated systems), probabilistic logic as well as measurements and transient interactions with an environment.
Thus, quantum process refers to the formalism concerning the time evolution of open quantum systems.

Under certain reasonable assumptions~\cite{pechukas1994reduced}, most quantum processes can be described by the quantum operation formalism.
These are quantum dynamical maps, which are linear and completely positive (CT) map from the set of density matrices to itself.
Typically they are non-trace-increasing maps, and trace-preserving (TP) for quantum channels.
In quantum process tomography, the unknown quantum process is assumed to be a CPTP quantum operation.

Since the unitary has to act on both sides of the initial density matrix means it is a superoperator, i.e. an object that can act on operators like the state matrix.
Using a special matrix multiplication identity this can be written as $\ket{\rho}\rangle \rightarrow \mathcal{U}\ket{\rho}\rangle$, where, $\mathcal{U} = U^* \otimes U$ and $\ket{\rho}\rangle = vec(\rho)$.
This looks like the pure state case because the operator (the state) has become a vector and the superoperator (the left right action of $U$) has become an operator.

A quantum process is a linear superoperator which acts on an input density operator to produce an output density operator.
For a quantum system with an input state $\rho_{in}$ of dimension $n\times n$ and an output state $\rho_{out} = \mathcal{E}(\rho_{in})$ of dimension $m\times m$, we can view this system $\mathcal{E}$ as a linear map between the space
of Hermitian matrices $\mathcal{E} : \mathcal{M}_{n\times n} \rightarrow \mathcal{M}_{m\times m}$.
For the case of unitary operation this is an isomorphism.
While $\rho$ is an order 2 tensors (i.e. operators), acting on Hilbert spaces of dimension $D=2^n$, $\mathcal{E}$ is an order 4 tensor specified by $D^4-D^2$ parameters.
Beside the superoperator, there are other equivalent~\cite{de2019fault} representations of quantum processes like Choi-matrix $\Lambda$, Kraus operators, Stinespring, Pauli basis Chi-matrix $\chi$, Pauli Transfer Matrix, Lindbladian, etc.

Let $\{B_i\}$ is any orthogonal basis for $\mathcal{M}_{d\times d}$ such  that $\langle B_i,B_j \rangle = d\delta_{ij}$ and $d= 2^n$.
Often the Pauli basis is chosen for this.
The process map can be written in the form $\mathcal{E} = \sum_{m,n=0}^{d^2-1} \chi_{m,n} B_m \rho B_n^\dagger$.
The $d^2\times d^2$ matrix $\chi$ is called the process matrix and together with the basis $\{B_i\}$, completely characterizes the map.
Typically this is used for QPT using the standard method.
The disadvantage of this is that besides the $\chi$ matrix, the basis also needs to be specified as a free-parameter to uniquely describe the process.

In this research we would prefer the Choi representation.
The Choi matrix $\rho_{Choi}$ is the density matrix obtained after putting half of the maximally entangled state $\ket{\Omega}$ through the channel $\mathcal{E}$, while doing nothing on the other half.
Thus it requires $2n$ number of qubits, but since the input state is fixed, we effectively do a quantum state tomography (QST) on this larger space instead of QPT, reducing the overall number of trials.

$$\rho_{Choi} = (\Lambda \otimes I)(\ket{\Omega}\bra{\Omega}) = \sum_{i,j} \dfrac{1}{2^n} \Lambda (\ket{i}\bra{j})\otimes \ket{i}\bra{j}$$

As a result of the Choi-Jamiolkowski isomorphism, the Choi matrix $\Lambda $ characterizes the process $\mathcal{E}$ completely.
This forms the basis of the channel-state duality between the space of CP
maps and the space of density operators. 
The Choi matrix can be transformed from the $\chi$ matrix by converting between the Pauli basis and the Bell basis.

\subsection{Entanglement-assisted QPT}

To fully characterize any quantum state its density matrix needs to be determined. 
Provided many copies of the state can be prepared, these are used to measure the state in a full set of measurement observables. 
Typically the Pauli basis is used, with observables X, Y and Z for every qubit, resulting in a total of $3^n$ different measurements. 
By repeating, the expectation value of each observable can be determined, and from this the density matrix state can be reconstructed. 
This process is known as quantum state tomography.

When the action of a quantum mechanical operation on an arbitrary input state needs to specified completely, a description needs to be found that characterizes the operation as an open quantum system via quantum process tomography.
The most straightforward method is standard quantum process tomography.
In general QPT is very costly as there are at least $12^n$ different experiments to specify an n-qubit system, which all need to be performed a large number (1000-10000) of times to estimate the expectation values of those experiments.

The expectation value of a measurement observable $M$ on a qubit in the state $\rho$ is denoted as $E_{\rho,M} = \text{tr}[M\rho]$. 
If the expectation values of a full set (spans the operator space) of observables are known, the state can be specified using standard QST:
$$\rho_{est} = \sum_{P_k \in \mathcal{P}_n} \dfrac{1}{2^n} E_{\rho,P_k} P_k$$
where $\mathcal{P}_n$ is the Pauli basis in n-qubits.
A recent line of research tries to minimize the number of experiments based on classical shadow states~\cite{huang2020predicting}.

For SQPT the action of the operation on an informationally complete (IC) set of input states is characterizing by performing QST on all corresponding outputs.
The IC set is for which the density matrices within that set span the entire operator space and is minimally $4^n$, though it is easier to use the $6^n$ Pauli eigenstates. 
The expectation value of a state $\rho_{out} = \mathcal{E}(\rho_{in})$ is given by:
$$E_{\rho_{in},M} = \text{tr}[M\rho_{out}] = \text{tr}[M\mathcal{E}(\rho_{in})] = \text{tr}[M\sum_{m,n=0}^{d^2-1} \chi_{m,n} P_m \rho P_n] $$
This can be solved by a set of linear equations, that links the terms with the process matrix by using measurement estimates from a total of $18^n$ different experiments.

In EAPT instead, the system of interest is first connected to an ancillary system of (at least) the same size. 
The maximally entangled state $\ket{\Omega}$ is prepared on the $2n$ qubits and the process is applied to half of the system.
If instead the separable Werner state is used it is called AAPT.
This results in the Choi matrix density state on the composite space which can be estimated using QST.
Standard QST measuring all qubits in the 3 different bases, results in a total of $3^{2n} = 9^n$ different measurements.
In comparison to SQPT it is a trade-off between the number of qubits (space) and the number of measurements (time) of the tomographic reconstruction process.
This one-to-one map enables all of the theorems about quantum operations to be derived directly from those of quantum states as a quantum process can be formalized as a quantum state in a larger Hilbert space.
Thus, an agent capable of performing QST can also reduce QPT and characterize the process.







\subsection{QKSA guided QPT}

Given the generality of quantum processes for modeling classical and quantum dynamics, QPT is chosen as the core algorithm that the AGI agent QKSA tries to optimize.
The agent is given access to a black-box that it the environment it tries to characterize, model and predict.
From QPT, it is clear that if the process matrix is known, the prediction for any input state can be obtained.
Thus, QPT reconstruction is the learning/training phase of the agent.
Currently static environments are considered, but for dynamic environment the process matrix needs to be continuously updated.

The QKSA searches over various QPT strategies that are admissible given the bounds of the \textit{least} metrics.
From this pools of QPT algorithms, the QKSA chooses the algorithm that has the least cost based on the evolving cost function for the specific QKSA.
For example, bounding the memory/space could sieve out EATP due to the limitation of the number of qubits, whereas, bounding the run-time could sieve out SQTP due to the high number of iterations.
The approximation value would directly influence the number of trials for each experiment used for the estimation in QPT.
The overall algorithmic complexity of the QPT algorithm can also be easily estimated.

For each environmental interaction, the QKSA chooses a QPT strategy based on the cost function and the \textit{least} metric, as well as the specific input state and measurement basis as dictated by the chosen QPT algorithm.
Thus the QKSA simulates an autonomous quantum physicist characterizing a quantum process.


\section{Conclusion} \label{s8}

In this article we present the framework for a Quantum Knowledge Seeking Agent (QKSA).
It is an artificial general intelligence (AGI) model that interacts with the environment as a general reinforcement learning (GRL) strategy.
We assume universal computation for building an inductive model.
However, the computation is bounded on various resources like hypothesis length, energy cost, approximation, space/memory and run-time.
A cost function based on these estimates of these 5 parameters (called the \textit{least} metrics) evolves as a genetic program (GP) within the QKSA.
The GP creates a new QKSA with a mutated cost function when the prediction capability performs below a certain bound, and halts/dies for an even lower performance.
The reproduction is implemented by embedding the QKSA within a quine as a self-replicating artificial life to maintain the exploration-exploitation trade-off in GRL.
This idea of biosphere preserving old models which becomes useful in changing environments can also be applied to physical models (e.g. Newtonian physics is an approximation of relativity, and easier to use for most everyday scenarios).

The QKSA is applied to the task of quantum process tomography (QPT) where it learns the best reconstruction strategy given the computation resource cost and bounds.
Thus, the QKSA is an active (participatory)~\cite{grinbaum2013quantum,fields2018sciences} version of the prescription~\cite{mueller2020law} to reconstruction quantum physics from Solomonoff induction using observer states, in line with John Wheeler's vision of a ``Law without law".

\subsection{Implementation}

We are currently implementing the model described in this article.
Developments in this direction can be accessed from the open-sourced repository at \href{https://github.com/Advanced-Research-Centre/QKSA}{https://github.com/Advanced-Research-Centre/QKSA}.

The results from the implementation will be published subsequently.
In this article we focus on the overview of the formal model and the framework definition to lay the foundation for collaborative research that is being undertaken for QKSA.

\subsection{Future work}
 
Some of the directions that are being researched to extend the QKSA framework are briefed here:

\begin{itemize}[nolistsep,noitemsep]
    \item \textbf{Classical physical environment}: classical physics models can be built with QKSA as it is a subset of quantum information. Simple models like grid navigation or game environments from OpenAI Gym Universe can be explored and compared with agents like AIXI.
    \item \textbf{Classical games}: winning strategy for classical turn-based games like tic-tac-toe can be learnt. Given the board state the agent chooses the optimal action for the current turn based on strategies that trades off computation time with memory or other resources (e.g. in card matching memory game).
    \item \textbf{Multi-threaded quines}: the physical implementation of quines need to share the resource of the underlying simulation hardware. A hypervisor to execute and gain insight of the evolutionary branches of the quines is being developed for the QKSA.
    \item \textbf{Interacting multiple agents}: in the current quine model the environment is replicated for each agent. However, as an extension, the QKSA can be built on a single environment with embedded agents. In such scenarios, agents can interact via the environment, without any special communication channel. Thus, in a noisy environment, the inter-agent communication would also be affected.
    \item \textbf{Collaborative modeling}: in the multi-QKSA extension, the reason for communication can be to communicate the learnt models, thus, effectively increasing the total perceptions from the environment and giving a boost to the agent's modeling process. In this case we will typically consider cooperative agents (non-adversarial collaborators). In this capacity it can explore thought experiments like Wigner's friend and the Frauchiger-Renner extended scenario, as well as models of relativistic QM and QBism.
    \item \textbf{Quantum acceleration}: the computation within the QKSA can be boosted with quantum acceleration as part of the agent, or as an environmental resource. 
    \item \textbf{Comparison with neural networks}: RL strategies based on neural networks like RNN/LSTM and RBM are already employed for QPT. While QKSA is more general, it would be worthwhile to compare between these models. 
    \item \textbf{Quantum cellular automata environment}: the environment can be modeled as a quantum cellular automata, i.e. local unitary evolution with (entanglement and) superposition, to model physical k-local Hamiltonians. Since these are universal~\cite{t2016cellular}, QCA can form a restrictive yet expressive subset for QKSA.
    \item \textbf{Causal structure discovery and modeling}: the QKSA can be used as a general technique for causal modeling in Bayesian inference, formal model checking, efficient digital twins and explainable AI research.
    \item \textbf{Variational optimizer}: the QKSA can model the currently popular hybrid quantum-classical heuristic algorithms based on the variational principle. In that setting, the QKSA would be a RL based classical optimizer for the quantum ansatz. 
\end{itemize}

While the list above is broad, it is worthwhile to mention some of the direction that QKSA currently steers clear of.
QKSA is based on an epistemic modeling of physics based on a RL based AGI.
It uses GP and Quines to evolve the AGI fitness function.
However, though this shares many features, we do not currently consider QKSA as a quantum artificial life~\cite{alvarez2018quantum} model as the agent is classical.
As an extension, QKSA is not related to quantum biology research.
QKSA also shares the limitations of \cite{mueller2020law}.
This means that currently the focus is on non-relativistic models of classical and quantum dynamics, and would not help to learn gravitational models.
We strongly believe the holographic duality of AdS/CFT~\cite{li2019measuring} would help us discover an unified theory based on quantum information and computation principles.








\bibliographystyle{unsrt}
\bibliography{ref.bib}

\end{document}